\documentclass[prb,showpacs,reprint,superscriptaddress,aps]{revtex4-1}
\usepackage{amssymb}
\usepackage{amsmath}
\usepackage{graphicx}
\usepackage[]{hyperref}
\hypersetup{pdftitle={Electronic properties of graphene nanoribbons under gate electric fields},
pdfauthor={Tobias Burnus, Gustav Bihlmayer, Daniel Wortmann, Yuriy Mokrousov, Stefan B{\"u}gel, and Klaus Michael Indlekofer}}

\newcommand{\dd}{\mathrm{d}}
\newcommand{\ee}{\mathrm{e}}
\newcommand{\ii}{\mathrm{i}}
\newcommand{\Gpar}{{\mathbf{G}_\parallel}}
\newcommand{\Gperp}{{G_\perp}}

\begin{document}

\title{Electronic properties of graphene nanoribbons under gate electric fields}

\author{Tobias Burnus}
\affiliation{Peter Gr\"unberg Institut and Institute for Advanced Simulation,
Forschungszentrum J\"ulich, and J\"ulich Aachen Research Alliance, 52425 J\"ulich, Germany}

\author{Gustav Bihlmayer}
\affiliation{Peter Gr\"unberg Institut and Institute for Advanced Simulation,
Forschungszentrum J\"ulich, and J\"ulich Aachen Research Alliance, 52425 J\"ulich, Germany}

\author{Daniel Wortmann}
\affiliation{Peter Gr\"unberg Institut and Institute for Advanced Simulation,
Forschungszentrum J\"ulich, and J\"ulich Aachen Research Alliance, 52425 J\"ulich, Germany}

\author{Yuriy Mokrousov}
\affiliation{Peter Gr\"unberg Institut and Institute for Advanced Simulation,
Forschungszentrum J\"ulich, and J\"ulich Aachen Research Alliance, 52425 J\"ulich, Germany}

\author{Stefan Bl\"ugel}
\affiliation{Peter Gr\"unberg Institut and Institute for Advanced Simulation,
Forschungszentrum J\"ulich, and J\"ulich Aachen Research Alliance, 52425 J\"ulich, Germany}

\author{Klaus Michael Indlekofer}
\affiliation{Hochschule RheinMain, Am Br\"uckweg 26, 65428 R\"usselsheim, Germany}

\date{\today}


%
%


\begin{abstract}
Quantum-dot states in graphene nanoribbons (GNR) were calculated using density-functional theory, considering the effect of the electric field of gate electrodes. The field is parallel to the GNR plane and was generated by an inhomogeneous charge sheet placed atop the ribbon. Varying the electric field allowed to observe the development of the GNR states and the formation of localized, quantum-dot-like states in the band gap. The calculation has been performed for armchair GNRs and for armchair ribbons with a zigzag section. For the armchair GNR a static dielectric constant of $\varepsilon\approx 4$ could be determined.
\end{abstract}

\maketitle

\section{Introduction}

After their experimental realization, single and bilayer graphene sheets and nanoribbons have attracted intensive attention due to their peculiar properties, which make graphene and its derivatives one of the most prominent material classes for future nanoelectronics.\cite{Geim2007,Neto2009,Geim2009}
The vast range of possible applications is due to the high carrier mobility,\cite{Novoselov2005,Zhang2005,Avouris2007,DasSarma2011} and remarkably long spin lifetimes and phase coherence lengths, which are particularly valuable for quantum information processing.\cite{Trauzettel2007} The peculiar electronic structure of gapless semiconductor graphene prevents electrostatic confinement due to Klein tunneling and gives rise to unique transport properties.\cite{Katsnelson2005} Experimentally, graphene-based tunable nanodevices can be realized, as shown, e.g., for graphene nanoribbons,\cite{Chen2007,Han2007,Li2008} interference devices,\cite{Miao2007,Russo2008} and graphene quantum dots.\cite{Stampfer2008a,Ponomarenko2008} Graphene nanoribbons (GNRs) play a particularly important role since
they are often used in field-effect transistors (FET) setups where the current flow is regulated by a gate electric field.  Additionally, gated nanoribbons can also be used to create quantum dots. Hereby, the spin qubits in quantum dots are a promising candidate for quantum information processing,\cite{Loss1998} for which  graphene is better suited than III--IV semiconductors due to its reduced hyperfine coupling and spin-orbit interaction.\cite{Trauzettel2007,Kane2005,Min2006,HuertasHernando2006,Stampfer2008,Liu2009} For nanoelectronics applications, the preferred width of a GNR is in the range of 1--3 nm as for wider ribbons the band gap diminishes.\cite{Son2006,Fiori2007}

For a proper theoretical description of GNR FET and GNR-based quantum dots, both the effect of an external electric field and the precise electronic structures of the ribbons have to be considered.
In ribbons of small width the details of the termination and exact structure of the edges play an important
role for ribbon's electronic structure due to the quantum confinement effects, which have to be
properly taken into account. The two most common types of edge termination of the ribbons lead to very different electronic structures in the vicinity of the Fermi energy: while armchair ribbons are insulating with a band gap of up to approximately 2.5 eV, zigzag ribbons exhibit metallic edge states.\cite{Fujita11996,Wang2007} For the former, the band gap decreases with increasing width of the ribbon and oscillates in magnitude with a periodicity of three as more rows of carbon atoms are added.\cite{Son2006}  Experimentally, armchair ribbons with the width of 1~nm with well-defined edges\cite{Cai2010} and FET with widths down to 2 nm have been realized.\cite{Wang2008} On the other hand, ribbons with the width of around 50 nm have been used to create quantum dots.\cite{Ponomarenko2008,Guttinger2010}

The sensitivity of the electronic properties of narrow GNRs to an external electric field and to the details of the structure, as well as the mechanism for the formation of the quantum dot states, calls for a description and analysis within the highly accurate density-functional theory techniques (DFT). In this work, we study the properties of narrow GNRs in an electric field, generated by finite gates, by employing a new scheme specifically designed for this purpose and realized within the full-potential linearized augmented plane-wave (FLAPW) method. In contrast to common approaches, the latter scheme allows to consider any given distribution of charge in the gates, thus enabling the versatility necessary for studies of complex nanostructured devices from first principles. Here, we describe the details of this approach as implemented within the FLAPW
method,\cite{Krakauer1979} and study the development of the electronic structure and screening properties of narrow GNRs as the
strength of the electric field is varied. We focus particularly on the emergence of the quantum-dot like states,
localized under the gates, and analyze their spatial distribution and symmetry properties.



\section{Inclusion of the electric field}

In a realistic device the electronic properties of the graphene nanoribbon will be modified by gate electrodes. While more complex interactions
between the gates and the graphene nanoribbon exist, the most prominent effect of the gates is the electric field they generate.
These electric-field effects can be included in a calculation based on DFT,\cite{Hohenberg1964,Kohn1965,book:Fiolhais} and in
the past several different implementations of external fields in DFT codes have been reported.\cite{Bengtsson1999,Neugebauer1993,Meyer2001,[][{ (in German).}]phd:Erschbaumer,Erschbaumer1990,Heinze1999,Weinert2009} However, in contrast
to most of these approaches, which model a situation in which the electric field is applied normally to a surface, the key effect of the gates we study is the field and potential distribution in the plane of the graphene ribbon.

\subsection{Computational scheme}

We use the film FLAPW method\cite{Krakauer1979} as implemented in the \textsc{fleur} code\cite{fleur} for our calculations. Two slightly different schemes to include the electric fields
of the gates have been tested in which the effect of the gates
is modeled by: (i) a charged sheet far in the vacuum in which the charge varies parallel
to the GNR or (ii) a plane in the vacuum at which a varying boundary condition to the Coulomb potential is applied. While the second approach
is closer to the picture of a metallic gate applying an electric field, the generation of the potential from a charge distribution is performed in
all existing DFT codes and, hence, the first approach is easy to implement. We discuss all details of the implementation of the gate electric field
in the appendix.

For modeling the effect of a gate electrode on a GNR we have chosen a free-standing 13-carbon-atom-wide graphene nanoribbon of armchair type
[see Fig. \ref{fig:GNR}(a)]. The dangling bonds are passivated using hydrogen leading to a ribbon which is about 5.1 nm long and 1.7 nm wide,
 consisting of 312 carbon and 48 hydrogen atoms. The setup is periodically repeated in $x$-direction to form an infinite ribbon. As our code
requires a two-dimensional periodicity, we also have to repeat the ribbon in $y$ direction. A supercell approach with a separation of 5 {\AA}
between adjacent cells is used to simulate a single isolated GNR.
The center of the GNR is sandwiched between positively charged top and back gates whose charge is compensated by negatively charged top and
back gates such that both the GNR and the two charge sheets remain as a whole charge neutral. The positively charged gate has a size of
$25.6\times12.3$ \AA$^2$ and the distance between the GNR plane and the charge sheets is 7 \AA.
For simplicity, we used a homogeneous surface-charge-density distribution on the gates [method (i)]. However, for the dimensions of this system,
the potential at the position of the plane of the GNR is qualitatively the same for a gate with constant potential  -- the potential only
differs strongly close to the gate (not shown).
The calculations were performed with the Perdew-Burke-Ernzerhof (PBE) exchange-correlation functional.\cite{PBE}

\subsection{Screening of the electrical field}

\begin{figure}
\hbox{\vbox{%
  \vskip 0.6em\hbox{\small{(a)}}\vskip -1.5em\hbox{\hskip-0.8em%
  \includegraphics[width=.48\textwidth]{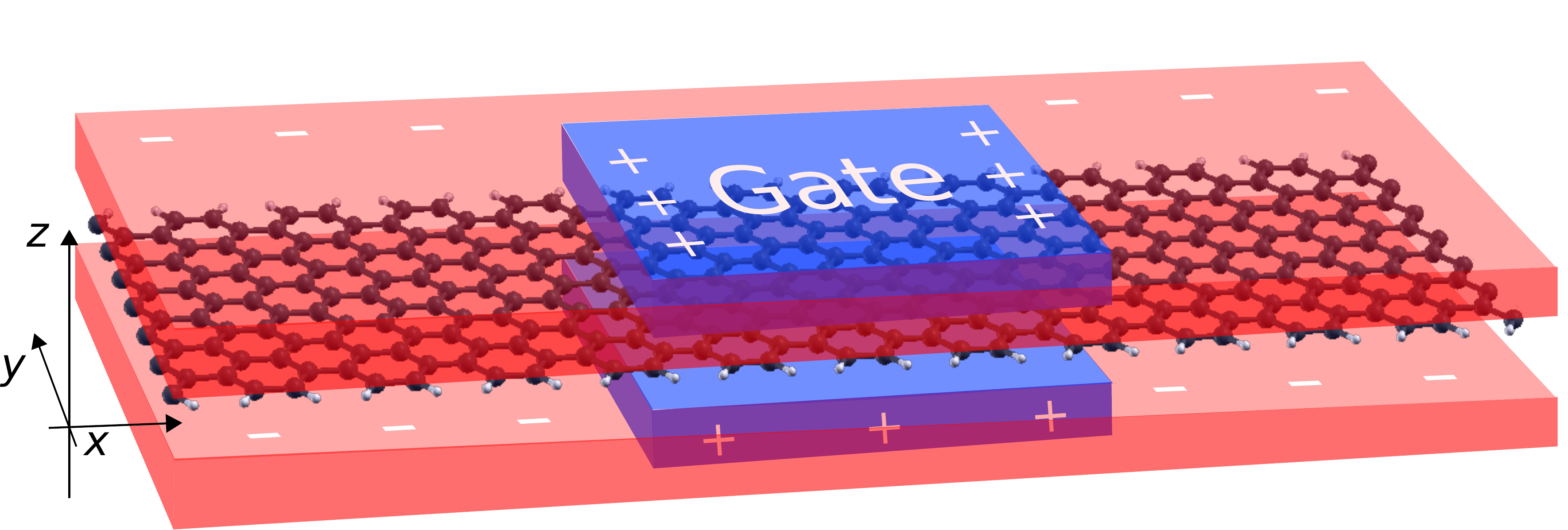}}}}
\hbox{\vbox{%
  \vskip 1.2em\hbox{\small{(b)}}\vskip -2.1em\hbox{\hskip-0.8em%
  \includegraphics[width=.48\textwidth]{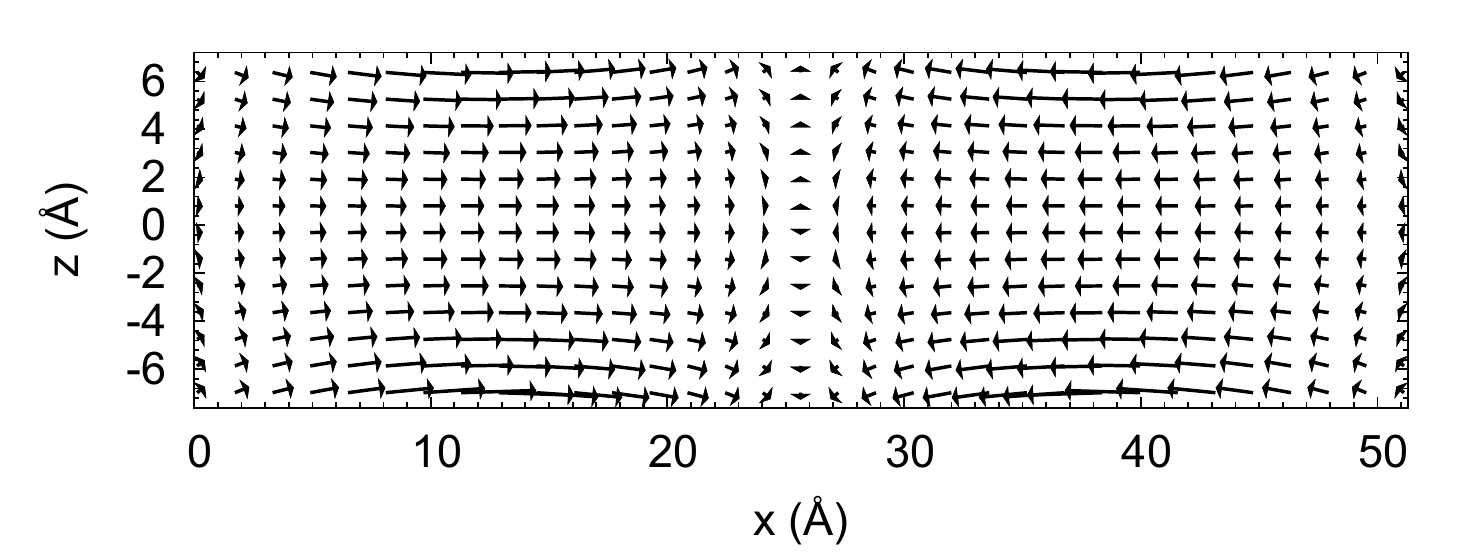}}}}
\caption{\label{fig:GNR} (Color online) (a) Free-standing, hydrogen-terminated armchair graphene nanoribbon of 13 atoms width. The blue rectangles indicate the location of the positively charged top and bottom gate electrodes. Their charge is compensated by the negatively charged
top and bottom electrodes (shown in red). (b) Side view showing the electric field created by the electrodes (for $y$ at the middle of the ribbon).}
\end{figure}

The height dependence ($z$ dependence) of the electric field is shown in Fig.~\ref{fig:GNR}(b), cutting vertically through the
GNR (with $y$ being at the middle of the ribbon).
In the plane of the ribbon ($z=0$), the field is fully in plane and parallel to the $x$ direction. Above the plane ($|z|>0$),
an out-of-plane component exists, which grows towards the center ($x=2.55$~\AA) and the left/right side ($x=0$ and $x=5.1$~\AA).
At exactly those points, the field is perpendicular to the GNR plane and the field strength is low, while for the points in between the field is
nearly in-plane and stronger. Thus, the potential should cause a charge accumulation under the gate.

\begin{figure}
\hbox to 0.47\textwidth {\strut\hfil\hbox{\vbox{
\vskip 0.7em\hbox{\small{(a)}}\vskip -1.7em\hbox{~~
\includegraphics[width=.43\textwidth]{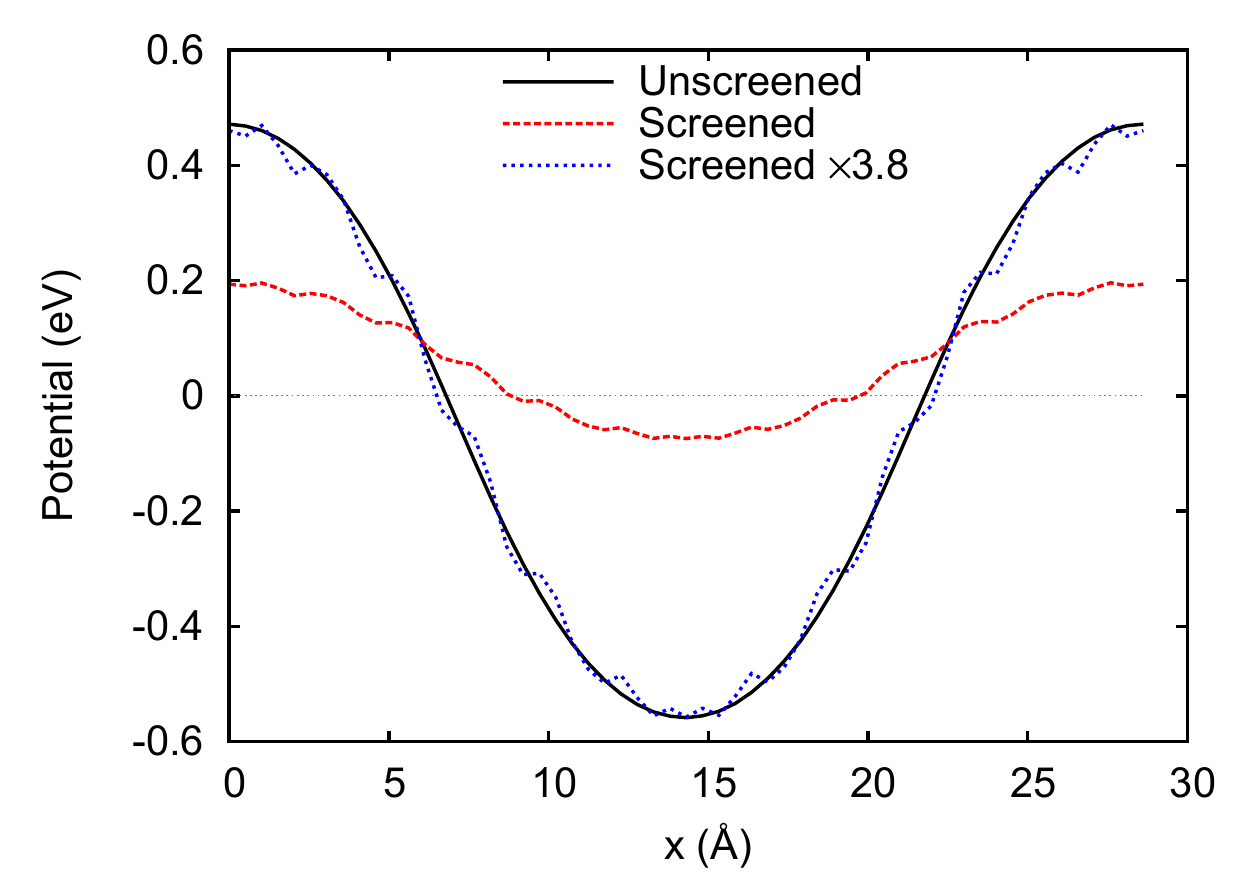}
}
\vskip 0.7em\hbox{\small{(b)}}\vskip -1.7em\hbox{~~
\includegraphics[width=.43\textwidth]{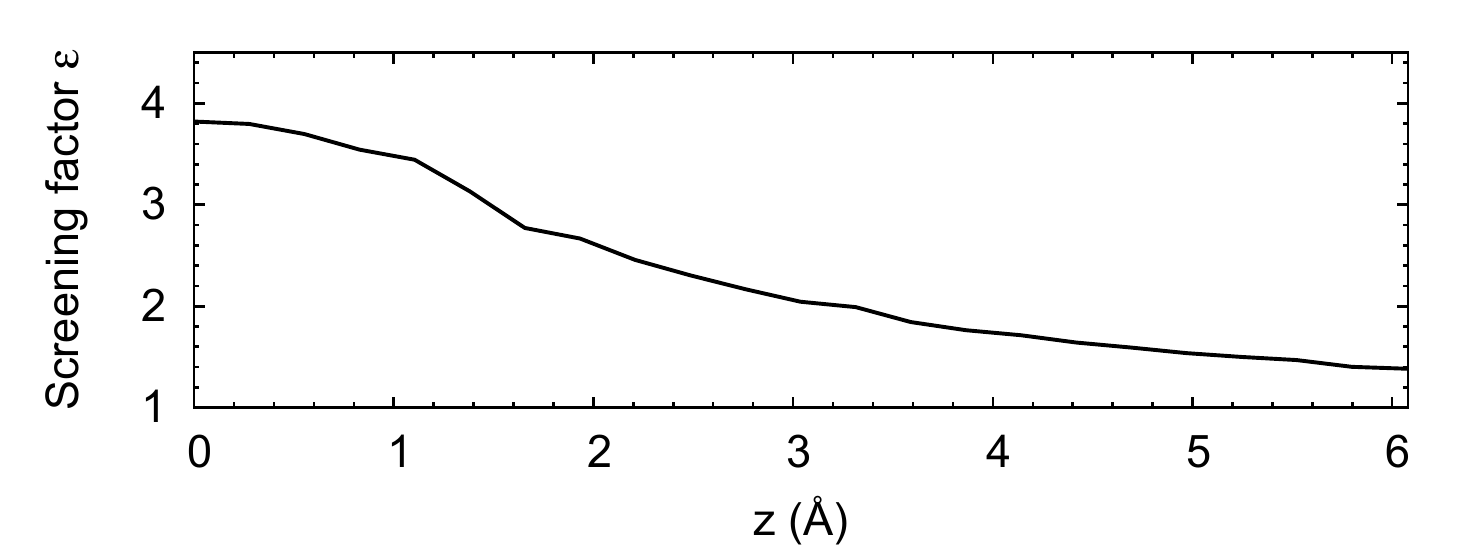}
}
}}\hfil}
\caption{\label{fig:screening} (Color online) (a) Potential due to the gate electric field along center of the GNR ($z=0$). The black solid curve shows the applied gate potential, the red dashed curve the screened potential. The blue dotted curve is obtained by  multiplying the red dashed curve by 3.8, indicating a static dielectric constant of $\varepsilon\approx 3.8$. (b) Dependence of the static screening on the distance from the plane.}
\end{figure}

\begin{figure}
\includegraphics[width=.48\textwidth]{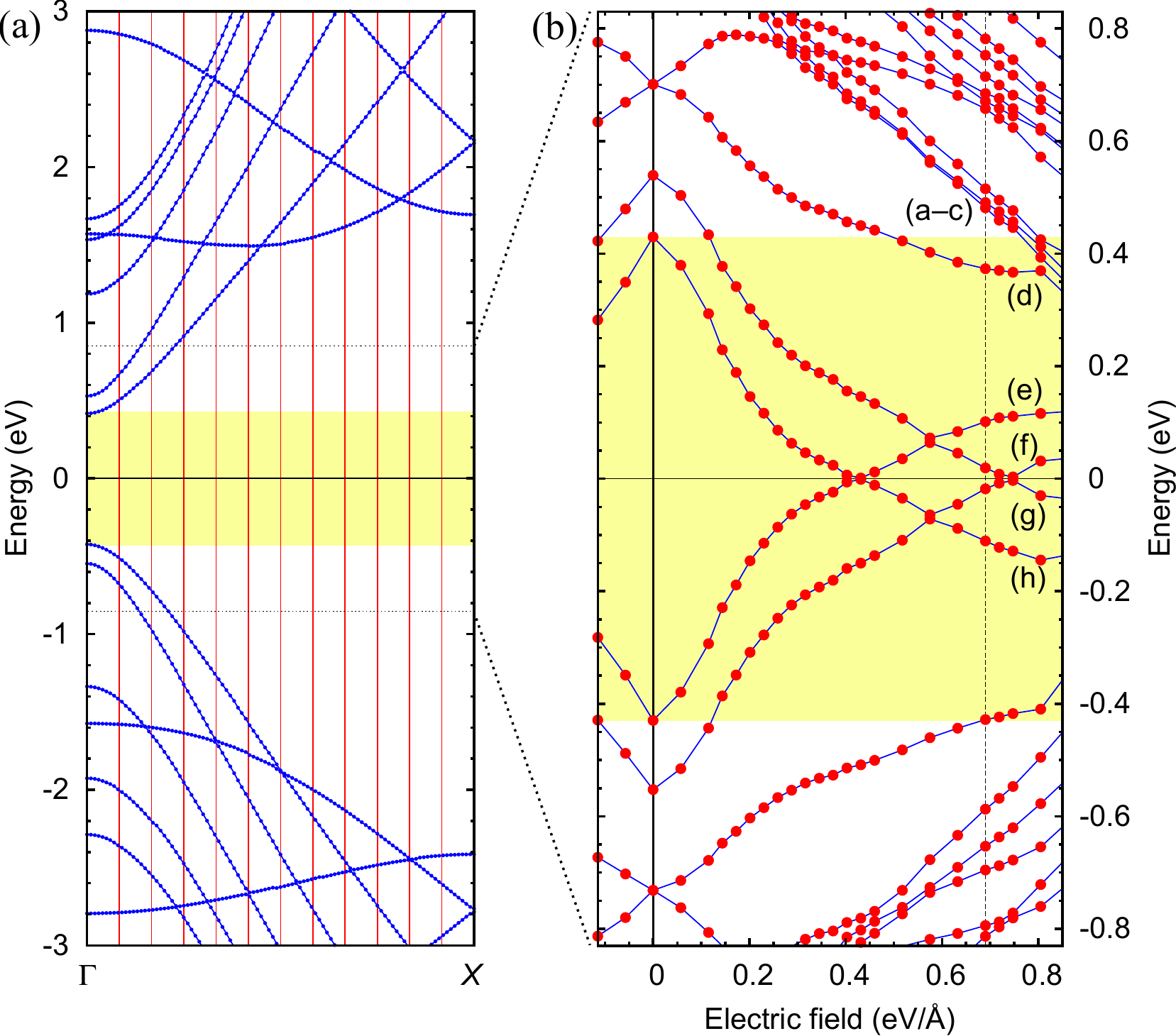}
\caption{\label{fig:eigen} (Color online) (a) Band structure for the smallest, 13 carbon atom wide GNR ($^1\!/\!_{12}$ as long as the gated GNR) with four staggered carbon atom in $x$ direction and totally 26 carbon atoms; the zero-field band gap of about 0.9 eV is shaded in yellow. (b) Single-particle eigenenergies gated GNR for the $\Gamma$ point in dependence of the applied electric field. The energy is relative to the Fermi energy; the abscissa shows maximal electric field (gradient of the potential before screening) at position of the ribbon. The states at the dashed vertical line [(a)--(h)] are depicted in Fig.~\ref{fig:den-single}.}
\end{figure}

Figure \ref{fig:screening}(a) shows the gate potential along the center of the ribbon for a gate charge of $0.0016 e/$nm$^2$.
These curves have been obtained by subtracting the total potential without an applied field from the potential with applied gate field canceling
the potential from the mostly inert ion cores.
The (in-plane) applied electric field at the position of the ribbon is given by the slope of the solid black curve and has
a maximal value of 0.056 eV/\AA.
The initial potential due to the applied electric field (black solid curve) gets reduced (red dashed curve) due to the screening
by the electrons in the ribbon. This screening effect is determined by evaluating the potential of the self-consistent calculation.
From the reduction -- the blue dotted curve has been obtained by multiplying the red dashed curve by 3.8 to match the black
solid curve -- a static ($\omega=0$) dielectric constant $\varepsilon\approx 3.8$ can be deduced for the center of the ribbon.
As the calculation has periodic boundary conditions (in $x$-direction) this static dielectric  constant $\varepsilon(q)$
corresponds to a wavevector of $q_0 = 0.123~\mathrm{\AA}^{-1}$.
The value of $\varepsilon(q_0,\omega=0)\approx 3.8$ agrees with the GW results of
Ref. \onlinecite{Schilfgaarde2011}. One should note, that the static dielectric constant has been evaluated
in the linear response limit of a small electric field, for stronger fields and
especially close to the crossover of states [cf. Fig.~\ref{fig:eigen}(b)],
different values are obtained.\cite{Burnus2011Eps}

The dependence of the screening on the distance from the GNR plane is shown in Fig.~\ref{fig:screening}(b);
$\varepsilon$ reduces only slightly from 3.8 to 3.4 over 1 \AA. This relatively constant screening
corresponds to presence of change density due to the $p_z$ orbitals, which is still significant at these distances.
For larger distances to the GNR $\varepsilon$ decreases exponentially, approaching the vacuum value of 1.

\begin{figure}
\includegraphics[width=.43\textwidth]{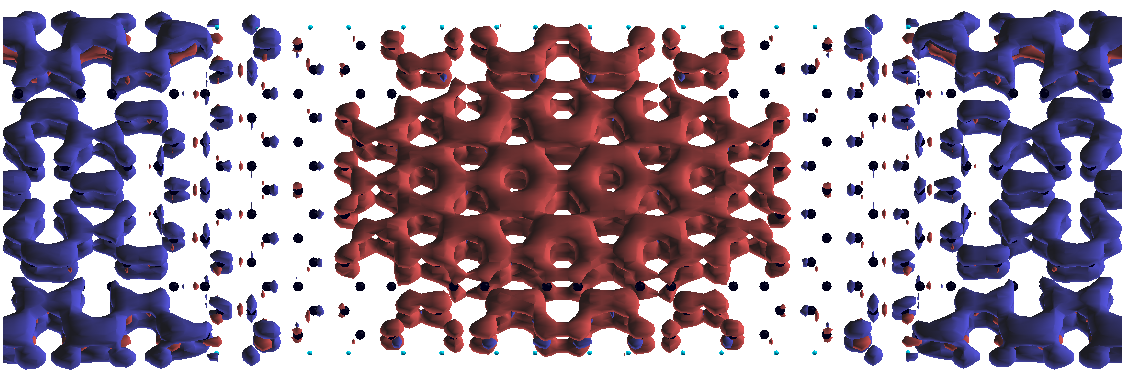}
\caption{\label{fig:total-den-diff} (Color online) Total charge density difference between a calculation with gate electric field and without; the density is reduced (blue) at outside and increased (red) in the center.}
\end{figure}

\begin{figure}
\hbox to 0.45\textwidth {\strut\hfil\hbox{\vbox{
\vskip 0.5em\hbox{\small{(a)}}\vskip -1.5em\hbox{\phantom{\small{(g)}}
\includegraphics[width=.28\textwidth]{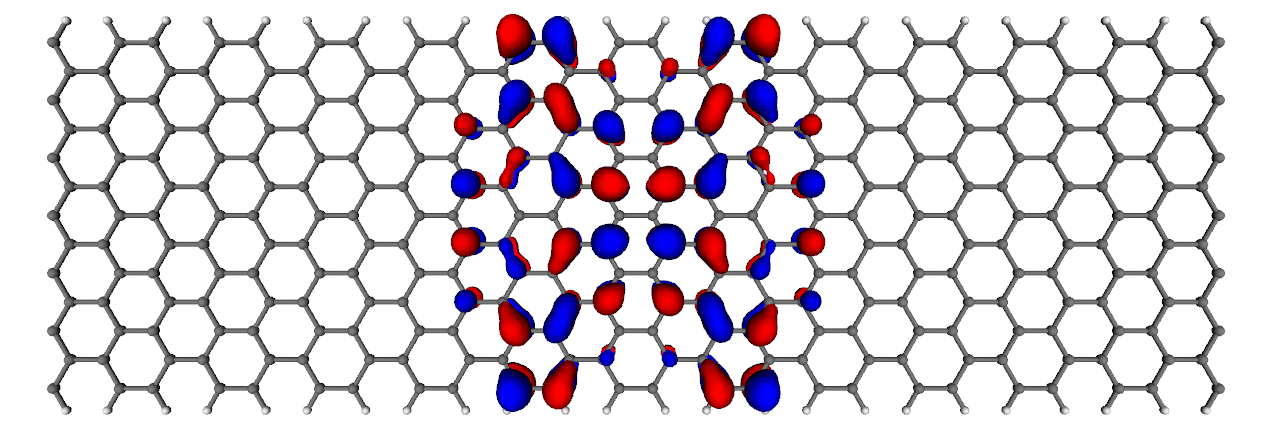}}
\vskip 0.5em\hbox{\small{(b)}}\vskip -1.5em\hbox{\phantom{\small{(g)}}
\includegraphics[width=.28\textwidth]{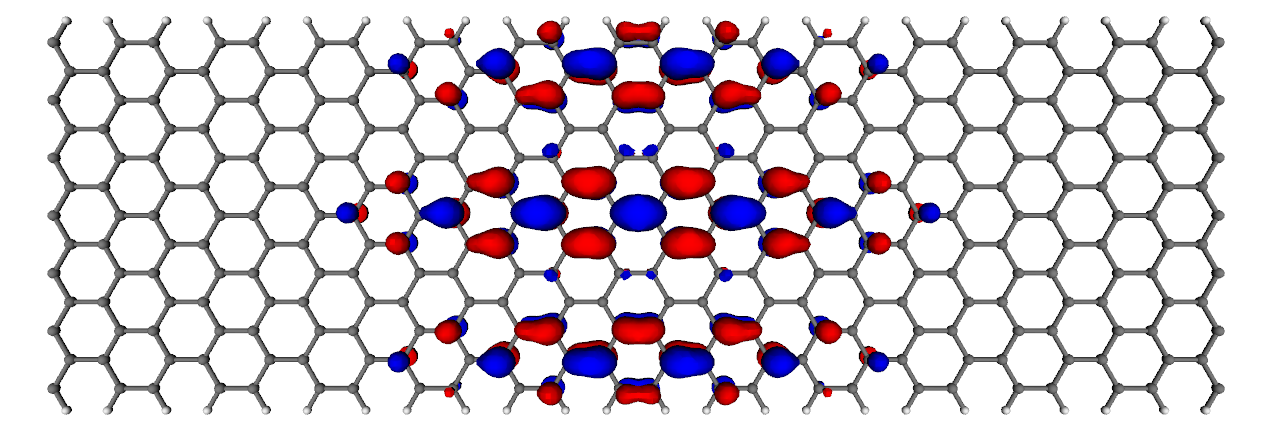}}
\vskip 0.5em\hbox{\small{(c)}}\vskip -1.5em\hbox{\phantom{\small{(g)}}
\includegraphics[width=.28\textwidth]{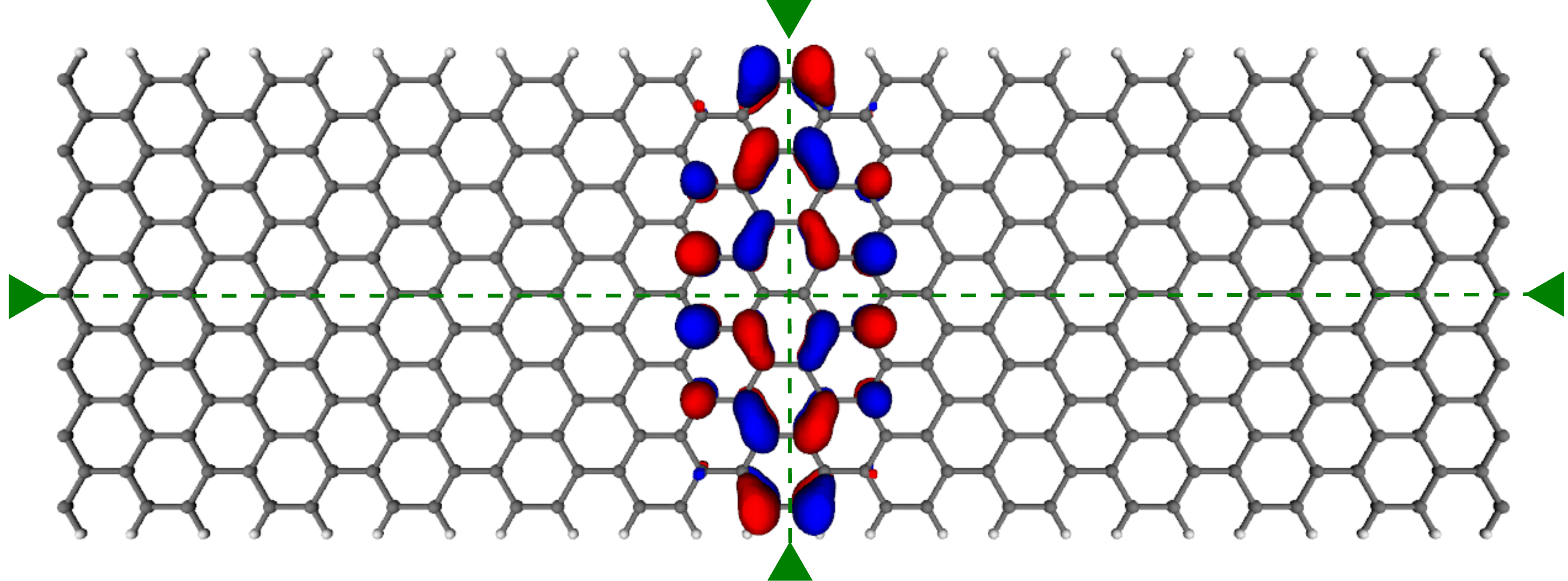}}
\vskip 0.5em\hbox{\small{(d)}}\vskip -1.5em\hbox{\phantom{\small{(g)}}
\includegraphics[width=.28\textwidth]{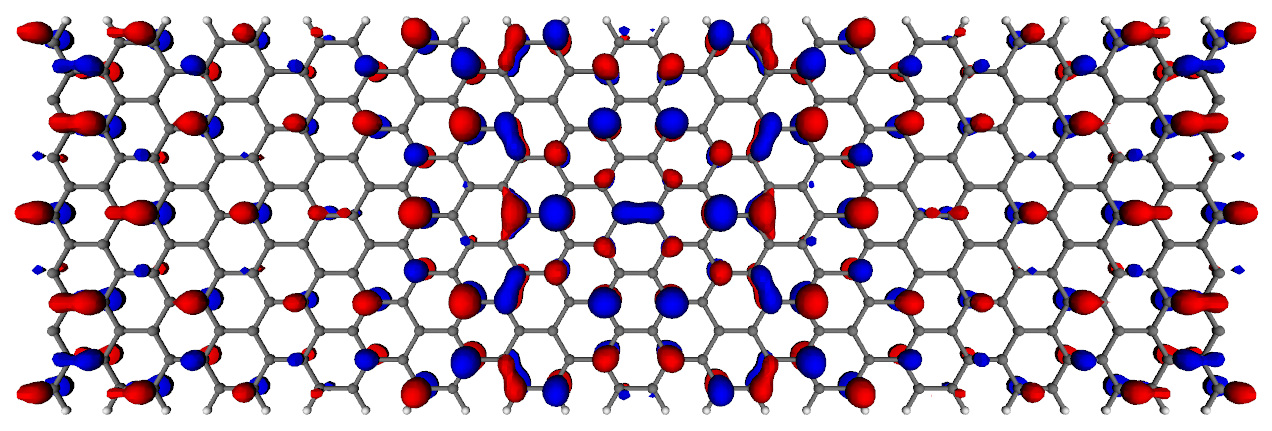}}
\vskip 0.5em\hbox{\small{(e)}}\vskip -1.5em\hbox{\phantom{\small{(g)}}
\includegraphics[width=.28\textwidth]{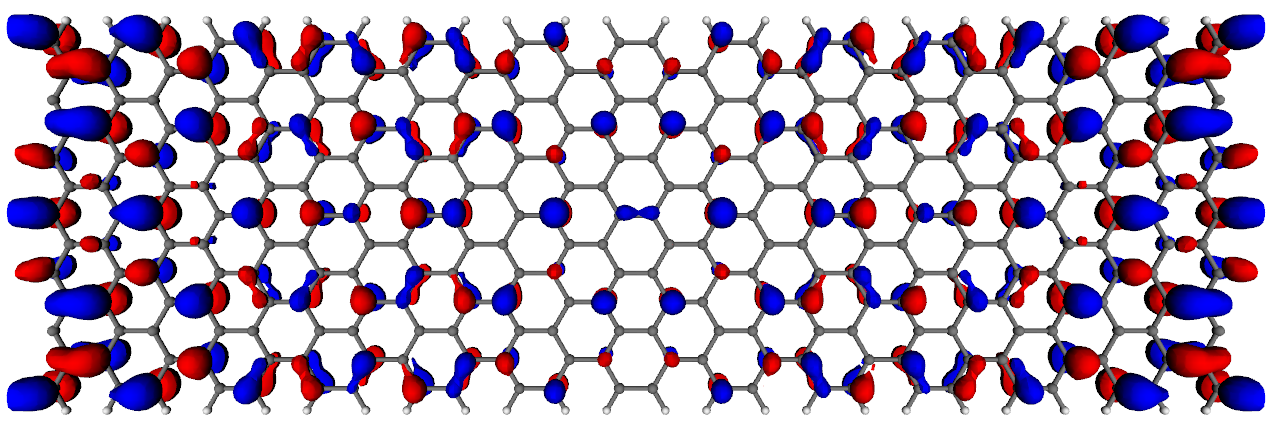}}
\vskip 0.5em\hbox{\small{(f)}}\vskip -1.5em\hbox{\phantom{\small{(g)}}
\includegraphics[width=.28\textwidth]{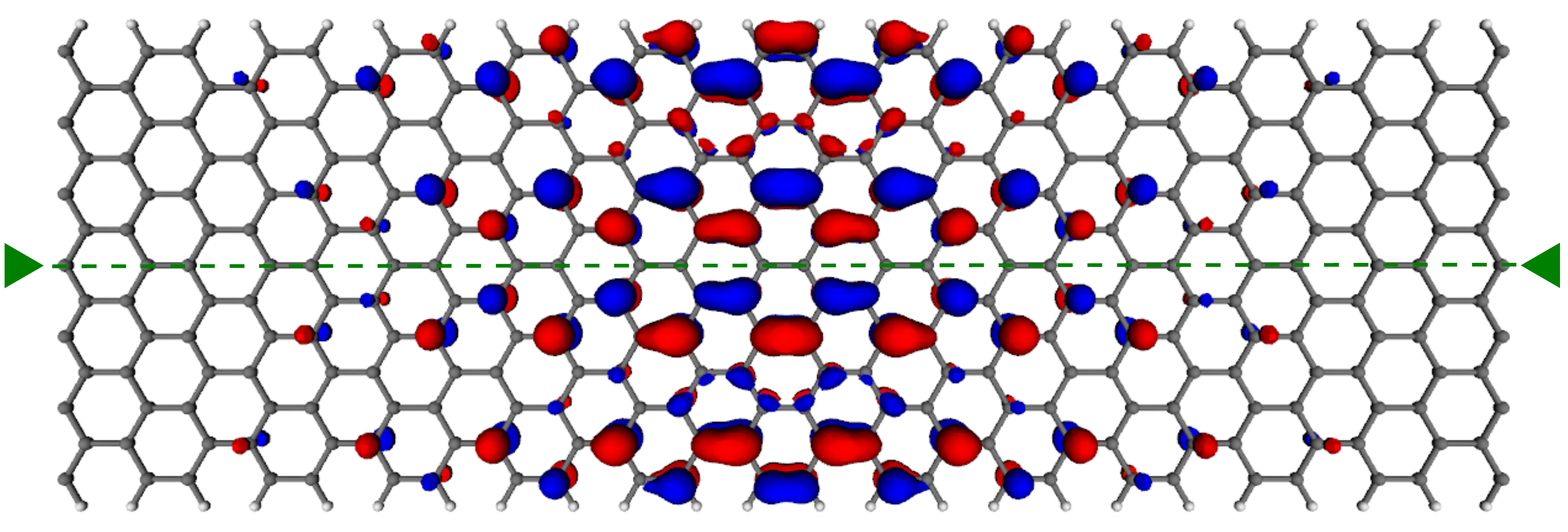}}
\vskip 0.5em\hbox{\small{(g)}}\vskip -1.5em\hbox{\phantom{\small{(g)}}
\includegraphics[width=.28\textwidth]{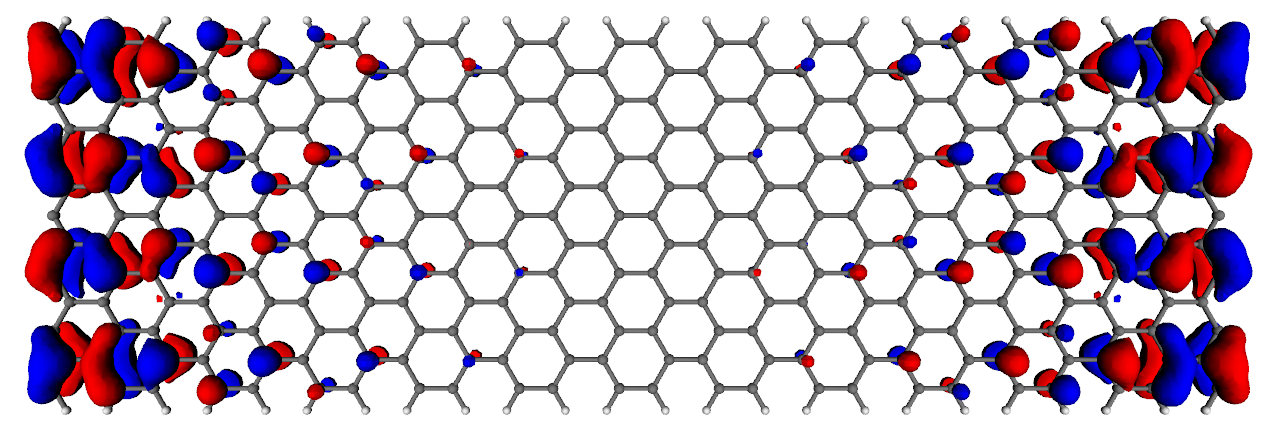}}
\vskip 0.5em\hbox{\small{(h)}}\vskip -1.5em\hbox{\phantom{\small{(g)}}
\includegraphics[width=.28\textwidth]{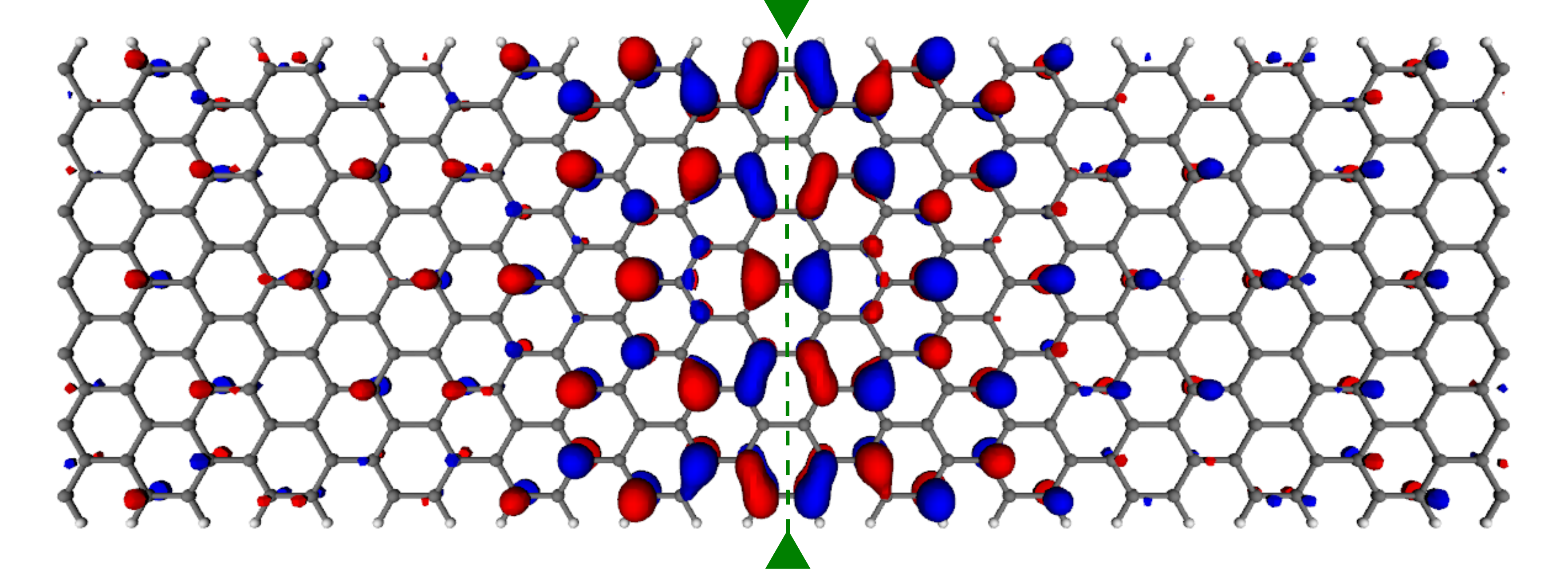}}
}}\hfil}
\caption{\label{fig:den-single} (Color online) Charge density of the single-particle levels at a field of 0.67 eV/\AA, ordered descending in energy; (a)--(f) are above, (g) and (h) below the Fermi energy. The color of the isosurfaces denotes the sign of the wavefunction.}
\end{figure}


\section{\label{sect:results}Results}

\subsection{Straight armchair ribbon}

In this section, we focus on the formation of quantum dot states in an armchair GNR with a width of 13 carbon atoms by applying an electric field as
outlined in the previous section. The edges of the ribbon are hydrogen-terminated and it is placed between the electrodes as visualized in
Fig.~\ref{fig:GNR}(a).  If one compares the total electron density of this structure without field and with a field of 0.67~V/\AA\ (this
corresponds to a charge of $\pm 0.19 e$/nm$^2$ on the gates), the charge accumulation and depletion can be visualized by a density-difference plot
as shown in Fig.~\ref{fig:total-den-diff}. Predominately charge of the $p_z$ orbitals is localized under the positively charged gate forming a
``quantum dot'' of about 20~\AA\ diameter.

To investigate the effect of the electric field on the electronic  structure of the GNR, we first consider in Fig.~\ref{fig:eigen}(a) the bandstructure
of the ribbon without field. The bandgap is about 0.9~eV and bounded by valence and conduction band states at the $\Gamma$ point. Introducing an
additional periodicity in $x$-direction by the field leads to a backfolding of the bandstructure along $\Gamma$--$X$. If the field is  modulated with
a periodicity of 51~\AA\ (12 unit cells in $x$-direction), the backfolding occurs at the red lines shown in Fig.~\ref{fig:eigen}(a). The Brillouin
zone extends then from $\Gamma$ to $X'$. In the following we focus on the states at the $\Gamma$ point, since they form the  valence and conduction
band edge.

Switching on the electric field leads to a splitting or a shift of the eigenenergies as shown in Fig.~\ref{fig:eigen}(b): The two-fold degeneracy
of the states at $-0.75$~eV and $+0.7$~eV, which results from the backfolding of the Brillouin zone, is lifted and the bands split almost symmetrically,
both for negative and positive fields. The slight asymmetry between positive and negative electric  field is introduced by the shape of the gate
electrode as shown in Fig.~\ref{fig:GNR}(a). If there were just a one-dimensionally modulated field acting on the GNR, this asymmetry would vanish.
The splitting can then be understood from the consideration of a system with just two eigenstates with energies $\epsilon_0^+$ and $\epsilon_0^-$,
perturbed by a periodic potential with Fourier coefficients $V_i$.  At the $i^{\rm th}$ backfolding of the bandstructure, the states split
according to $\epsilon_{1,2}(V) = \frac12 (\epsilon_0^+ + \epsilon_0^-)\pm\sqrt{\frac14(\epsilon_0^+-\epsilon_0^-)^2+|V_i|^2}$. As the states are
degenerate, the splitting is linear in $|V_i|$. In case of interacting states that are non-degenerate, the evolution of these states starts
parabolic at small fields and evolves into a more linear behavior at stronger fields ($V_i \gg (\epsilon_0^+-\epsilon_0^-)$).

With increasing field strength the uppermost valence band states  move up in energy, towards the Fermi level, while the lowest conduction band states
move down by a similar amount. Their interaction causes again a deviation from the linear behavior and
at a field strength corresponding to 0.4~V/\AA\ the first crossover of states occurs, i.e.\ a valence band state becomes unoccupied while a
conduction band state is populated. In order to get a better understanding of this scenario, we display the single-particle states in this
energy region in Fig. \ref{fig:den-single} for a field of 0.67 eV/\AA\ [vertical line in Fig.~\ref{fig:eigen}(b)].
The first crossover occurs for states shown in Fig.~\ref{fig:den-single}(e) and (h), i.e.\ with state (h) the first quantum dot state gets
localized under the gate electrode. State (e) has more density outside the positively gated region; it raises in energy with increasing field
and gets unoccupied.

At stronger fields (0.72 eV/\AA) the state Fig.~\ref{fig:den-single}(f) gets occupied, which can be regarded
as the second quantum dot state.  As for atoms, one expects also for quantum dots states an increasing number of nodal planes for higher
states with higher energy, which is indeed what one observes: The color of the isosurfaces in Fig.~\ref{fig:den-single} denotes the sign of the
wavefunction, which allows to locate the nodal planes.  The first quantum-dot state (h) has at the center one vertical nodal plane (dashed green
line) and is not nodeless as one might have expected. For the next localized state (f) one sees in the charge density a horizontal nodal plane, while
the state (c) has two: a horizontal and a vertical nodal plane. The state (b) seems to have two horizontal and the state (a) one horizontal and
two vertical nodal planes.

We have thus shown how under a gate electric field, localized states appear in the band gap; those show a nodal structure as one expects for a quantum dot. The states still show the structure of the underlying lattice and are predominately formed by $p_z$ states as one can see in the charge density.

\subsection{Z-shaped ribbon}

Besides armchair nanoribbons, also zigzag ribbons exist, which are known for their (conducting) edge states.\cite{Fujita11996,Nakada1996,Wang2007} Those are located just above and below the Fermi energy. The combination of both types of edge states could be useful for future nanodevices; but missing atoms or sections with different edge terminations might also occur involuntarily during the preparation of nanoribbons. We have thus simulated an armchair ribbon of width 13 (as above) but with a zigzag section in the middle.

\begin{figure}[t]
\vbox{
\vskip 2em\hbox{\small{(a)}}\vskip -3em\hbox{\phantom{\small{(a)}}
\includegraphics[width=.43\textwidth]{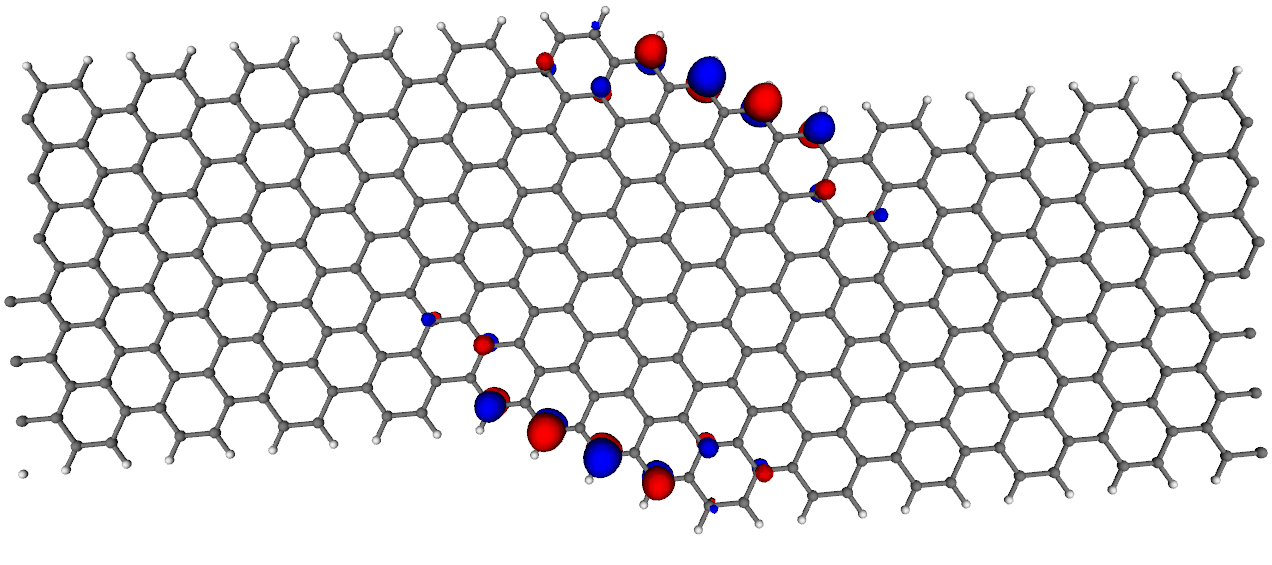}}
\vskip 2em\hbox{\small{(b)}}\vskip -3em\hbox{\phantom{\small{(b)}}
\includegraphics[width=.43\textwidth]{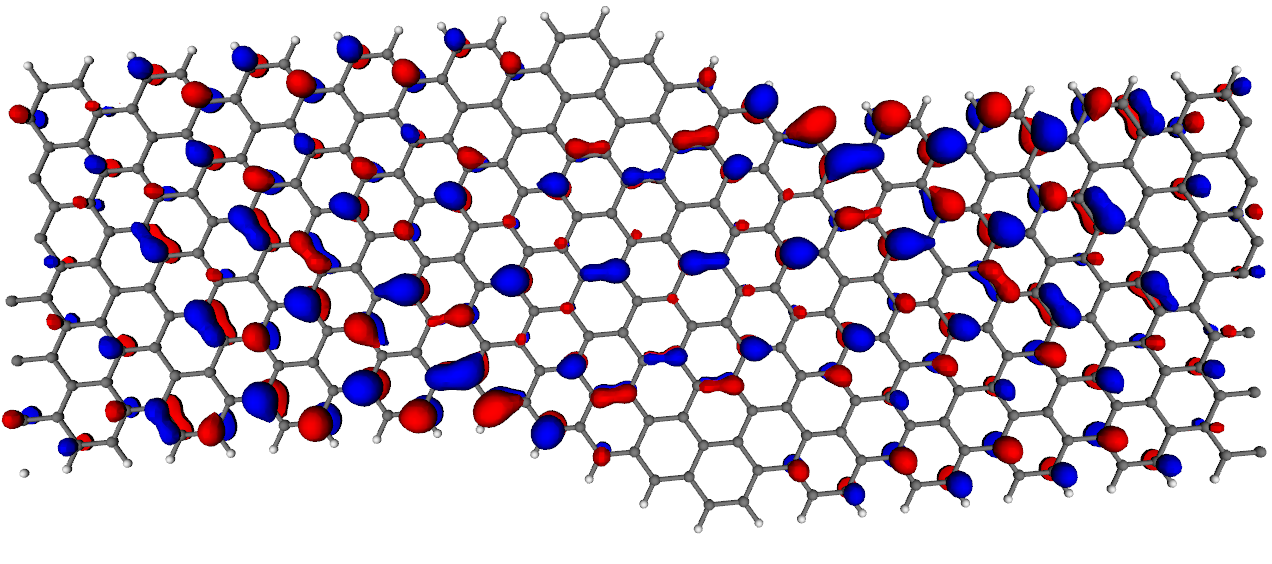}}
\vskip 2em\hbox{\small{(c)}}\vskip -3em\hbox{\phantom{\small{(c)}}
\includegraphics[width=.43\textwidth]{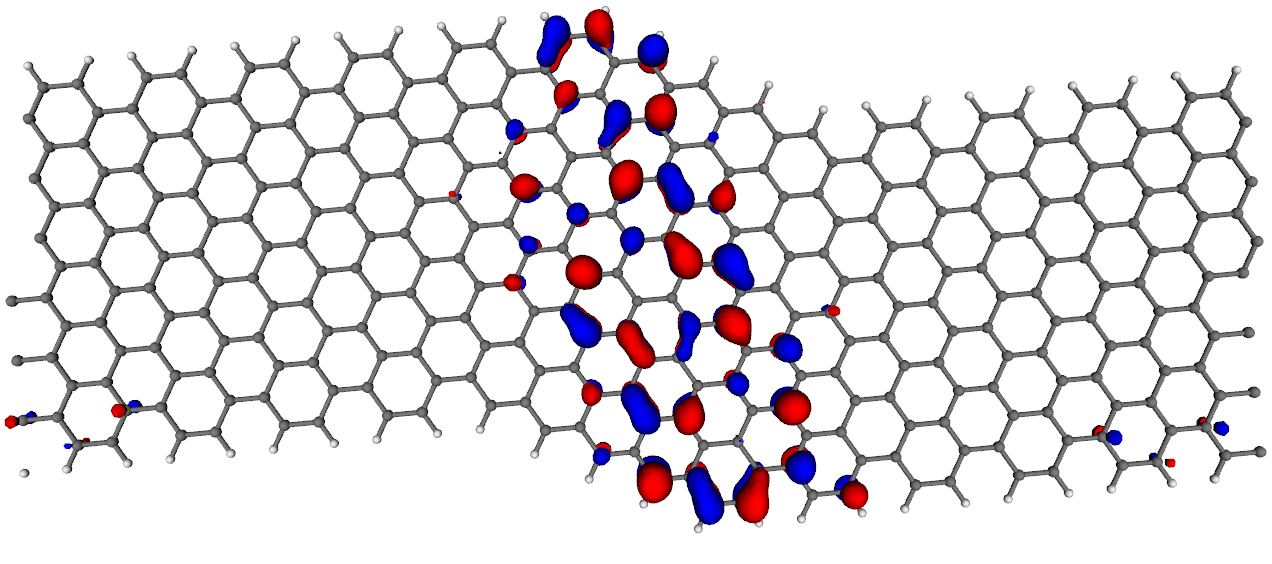}}
}
\caption{\label{fig:dens-zigzag} (Color online) Armchair nanoribbon with zigzag section. (a) Edge state at the Fermi energy (no field). (b) State above 0.87 eV above the Fermi energy (no field). (c): Same state with an applied gate field of 1 eV/\AA, now located at the Fermi energy.}
\end{figure}

For the calculation, a supercell with length of 51.3 \AA\ and width of 25.8 \AA\ has been used; the positive gate charge has been placed at the center over and below the zigzag region ($16\times18.8$ \AA$^2$). Figure \ref{fig:dens-zigzag}(a) shows (for zero field) the existence of an edge state in the zigzag section; the highest occupied (shown) and the lowest unoccupied state have the same density but locally different phases. If one now turns on an electric field, the edge states gain density in the middle of the ribbon. Additionally, states with more density at the outside raise in energy, while more localized states get more localized and become lower in energy. This is exemplified in Fig.~\ref{fig:dens-zigzag} for an electric field of 1 eV/\AA\ (maximal potential gradient before screening; gate charge is $0.026e$/nm$^2$): (b) shows a high lying state at zero field (0.87 eV above the Fermi energy), which moves down in energy to the Fermi energy in the electric field; at the same time its density becomes more localized at the zigzag region, i.e. under the gate.

\section{\label{sect:conclusion}Conclusion}

A scheme to include arbitrary electrical fields has been introduced and its implementation for the film FLAPW method outlined. The scheme allows for Neumann and Dirichlet boundary conditions and for differently shaped gate electrodes. Future use could encompass the calculation of the effects of an electrical field on adatom on films and a wide range other electric-field related phenomena.

The gate electric field was then applied to an armchair graphene nanoribbon; the mostly in-plane field caused a charge accumulation under the gate. For small fields, the linear response of the electrons in the ribbon to the electric field allowed to determine a static dielectric constant of $\varepsilon=3.8$. The field also drove states into the zero-field gap; as has been shown, the previously unoccupied states entering the gap region are localized under the gate. Those states, mostly formed by $p_z$ orbitals, show the structure of the underlying lattice. However, they also feature a nodal structure as one would expected for quantum dots. The states have been obtained from an all-electron density-functional theory calculation, which makes a comparison to less precise techniques such as tight binding interesting; those techniques have the advantage that much larger systems can be treated. Additionally, the influence of the electric field on GNR consisting of zigzag with metallic edge states and semiconducting armchair sections has been shown, where the electric field moves states towards the Fermi energy, which are localized under the gate.


\begin{acknowledgments}
The calculations were performed on the Juropa supercomputer at the Forschungszentrum J\"ulich, Germany.
The work was supported by Deutsche Forschungsgemeinschaft through Research Unit 912 ``Coherence and Relaxation Properties of Electron Spins". Y.~M. acknowledges funding under the HGF-YIG programme VH-NG-513.
\end{acknowledgments}


\appendix
\section{\label{sect:Appendix}Electric field in FLAPW}

\begin{figure}
\includegraphics[width=.43\textwidth]{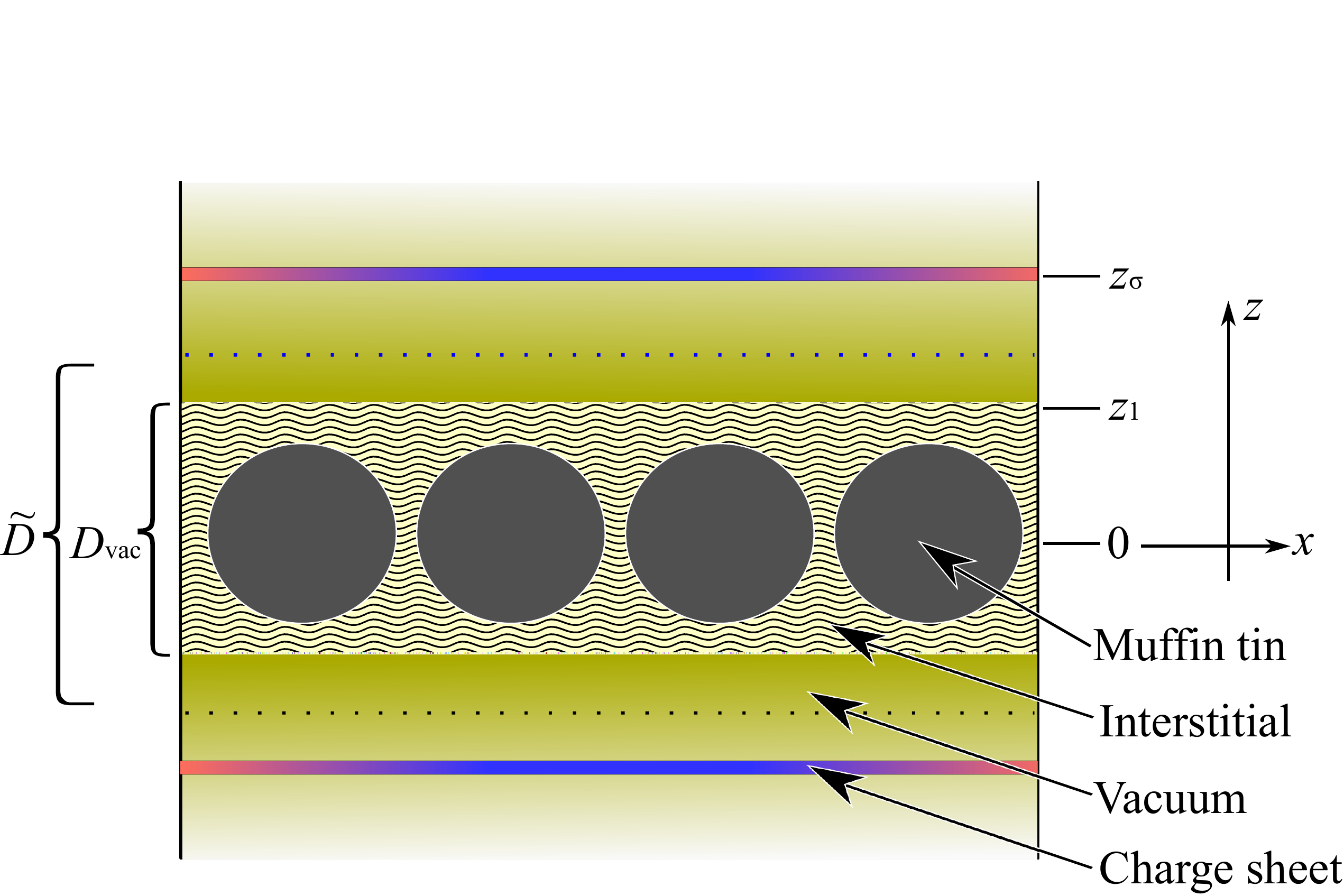}
\caption{\label{fig:film} (Color online) Geometric setup in the FLAPW method with muffin-tin spheres, interstitial, and two semi-infinite vacuum regions in $z$ direction. $|z|=z_1\equiv D_\textrm{vac}/2$ marks the interstitial--vacuum boundaries, $|z_\sigma|$ is the position for the charge sheets.}
\end{figure}

For periodic systems, the full-potential linearized augmented plane wave (FLAPW) basis, used for the calculations presented above, gives an accurate, all-electron description.\cite{Marcus1967,Koelling1975,Andersen1975,Krakauer1979,Wimmer1981,Weinert1982,book:Singh} Besides the more common FLAPW basis with three-dimensional periodicity, a film version of FLAPW exists. It has only two-dimensional periodicity and semi-infinite vacua in the third direction (chosen to be $z$).\cite{Krakauer1979,Wimmer1981,Weinert1982,[][ and references therein.]inbook:Bluegel2006,phd:Kurz} Hereby, the space is separated into three regions: the muffin-tin spheres around each atom, the interstitial region between the spheres and a vacuum region in which the density decays exponentially to zero; see Fig.~\ref{fig:film}. The basis set used for expansion of the wavefunction consists of the functions of the following form:

\begin{widetext}
\begin{eqnarray}\label{eq:basisfunc}
\varphi_{\Gpar \Gperp}(\mathbf{k}_\parallel,\mathbf{r})
=\begin{cases}\ee^{\ii(\Gpar+\mathbf{k}_\parallel)\mathbf{r}}
        \ee^{\ii \Gperp z},&
  \text{interstitial}\\
 \bigl[a_{\Gpar,\Gperp}(\mathbf{k}_\parallel)
       u_{\Gpar}(\mathbf{k}_\parallel, z)
   +   b_{\Gpar,\Gperp}(\mathbf{k}_\parallel)
       \dot u_{\Gpar}(\mathbf{k}_\parallel, z)\bigr]
   \ee^{\ii(\Gpar+\mathbf{k}_\parallel)\mathbf{r}} &
  \text{vacuum}\\
  \sum_L \bigl[a_L^{\mu \mathbf{G}}(\mathbf{k})u_l(r)
     + b_L^{\mu \mathbf{G}}(\mathbf{k})\dot u_l(r)\bigr]Y_L(\mathbf{\hat r}) &
  \text{$\mu$-th muffin tin}\end{cases},
\end{eqnarray}
\end{widetext}

\noindent where $\mathbf{G}=\Gpar+(0,0,\Gperp)$ is the reciprocal lattice vector, the $u_L$ are radial functions with $\dot u_L$ as their energy derivatives, $Y_L$ are spherical harmonics, $\hat{\mathbf{r}}=\mathbf{r}/|\mathbf{r}|$, and $a$ and $b$ are coefficients chosen such that the basis function is continuously differentiable across the interstitial--vacuum and interstitial--muffin-tin boundary. The $\parallel$ denotes the in-plane ($x$--$y$) and $\perp$ the out-of-plane ($z$) component. While the interstitial extends to $|z|\le z_1\equiv D_\mathrm{vac}/2$, the perpendicular wavevectors $\Gperp$ are defined with regard to $\tilde D > D_\mathrm{vac}$, i.e. $\Gperp = 2\pi n/\tilde D$, to allow for more variational freedom.

\subsection{\label{sect:Appendix:Neumann}Neumann boundary conditions}

The idea is to obtain the potential by integrating the surface charge density $\sigma$ from infinity, i.e.

\begin{eqnarray}\label{eq:VsigmaGeneral}
V(\mathbf{r}) = -\frac{1}{\varepsilon_0}\int_{-\infty}^{z}\sigma(\mathbf{r'})\,\dd z',
\end{eqnarray}

\noindent with $\dd\sigma/\dd z=\rho$, where $\rho$ is the charge density. The complication in evaluating the equation (\ref{eq:VsigmaGeneral}) arises from the use of different bases in the different regions. The electric field $\mathcal{E}$ could be included in the calculation via its associated potential $V_\text{EF}$ as an additional term to the external potential $V_\text{ext}$; however, in the presented scheme it enters as charge density $\rho_\textrm{EF}$. Their relation is given by $\varepsilon_0\nabla^2 V_\text{EF} =\varepsilon_0\nabla\cdot\mathcal{E} = -\rho_\text{EF}$, where $\varepsilon_0$ is the electric constant (for cgs replace $\varepsilon_0$ by $1/4\pi$).

For the vacuum region, the charge $\rho$ and the potential $V$ can be expanded in a Fourier series,\cite{[][{ (in German).}]phd:Erschbaumer,Erschbaumer1990,Weinert2009}

\begin{eqnarray}\label{eq:fourier}
V(\mathbf{r}) = \sum_{\Gpar}V_{\Gpar}(z)\ee^{\ii\Gpar\mathbf{r}},~
\rho(\mathbf{r}) = \sum_{\Gpar}\rho_{\Gpar}(z)\ee^{\ii\Gpar\mathbf{r}};\quad\quad
\end{eqnarray}

\noindent thus, the Laplace operator of the Poisson equation $\varepsilon_0\nabla^2V=-\rho$ separates into a $z$ and $\Gpar$ term,

\begin{eqnarray}\label{eq:poisson}
\varepsilon_0\left[\frac{\partial^2}{\partial z^2} - \Gpar^2\right]V_{\Gpar}(z) = -\rho_{\Gpar}(z),
\end{eqnarray}

\noindent which leads to two equations. For $\Gpar=0$ (uniform field in the vacuum), Eq.~(\ref{eq:poisson}) can be solved by integration. Using the boundary condition that the potential smoothly vanishes at infinity, $\lim_{z\to\infty}V(z)=0$ and $\lim_{z'\to\infty}\partial_zV(z)|_{z'}=0$, the potential in the vacuum is given by

\begin{eqnarray}
V_{\Gpar=0}(z) &=& -\frac{1}{\varepsilon_0}\int_z^\infty\!\!\!\!\int_{z'}^\infty\rho(z'')\,\dd z''\,\dd z'\nonumber\\
               &=& \frac{1}{\varepsilon_0}\int_z^\infty (z'-z)\rho(z'')\,\dd z'.
\end{eqnarray}

For the nonuniform part, $\Gpar\ne0$, the Green function

\begin{eqnarray}\label{eq:GreenF:Neumann}
G(z,z')=\frac{1}{2|\Gpar|}\ee^{-|\Gpar|\cdot|z-z'|}
\end{eqnarray}

\noindent can be used to give a particular solution for the potential,

\begin{eqnarray}\label{eq:gne0:V}
V_{\Gpar}(z) = \frac{1}{\varepsilon_0}\int_{z_0}^zG(z,z')\rho_\Gpar(z')\,\dd z',
\end{eqnarray}

\noindent which fulfills the boundary condition that the potential vanishes for $z_0\to\infty$ smoothly at infinity.

If the system has no mirror symmetry, e.g. because the field on the top is different from the one at the bottom, the charge density can be written as\cite{phd:Erschbaumer,WeinertUnpub}

\begin{eqnarray}
\rho^{(1)}(z) &=& \begin{cases}\displaystyle
\rho_0(z)+\bar\rho+\sum_{\mathbf{G}\ne0}\rho_\mathbf{G}\ee^{\ii \mathbf{G\cdot r}}, &  |z|\le z_1\\
0, & |z|>z_1
\end{cases}\\
\rho^{(2)}(z) &=& \begin{cases}
-\bar\rho, & |z|\le z_1\\ \displaystyle
\rho_0(z)+\sum_{\mathbf{G}\ne0}\rho_{\mathbf G}\ee^{\ii \mathbf{G\cdot r}},& |z|>z_1
\end{cases}
\end{eqnarray}

\noindent where $-\bar\rho$ denotes the average pseudo charge density of the interstitial, i.e.

\begin{eqnarray}
\bar\rho &=&\frac{1}{2z_1}\int_{-z_1}^{z_1}\rho_{\Gpar=0}(z)\,\dd z\nonumber\\
&=& -\rho_{0,0}-\sum_{\Gperp\ne0}\rho_{0,\Gperp}j_0(\Gperp z_1),
\end{eqnarray}

\noindent with $\rho_{\Gpar=0}(z)=\sum_{\Gperp}\rho_{0,\Gperp}\exp(\ii\Gperp z)$ and Bessel function $j_0(z)=\sin z/z$. To obtain the potential, one sums up the surface charge densities starting from minus infinity; the electric field is included as surface charge density $\sigma_{\rm EF}$. For the vacuum region, the $\Gpar=0$ component of the potential is given by

\begin{eqnarray}
V(z < -z_1) &=& -\frac{1}{\varepsilon_0}\int_{-\infty}^z\sigma(z')\,\dd z'\nonumber\\
            &&  - \theta(z-z_\sigma^{(-)})\frac{1}{\varepsilon_0}\int_{z_\sigma}^z\sigma_\textrm{EF}^{(-)}\,\dd z'\nonumber\\
V(z > z_1) &=& -\frac{1}{\varepsilon_0}\int_{z_1}^z \left[\sigma(z')+(-\bar\rho) D_\textrm{vac}+\sigma(-z_1)\right]\,\dd z'\nonumber\\
&&           - \theta(z-z_\sigma^{(+)})\frac{1}{\varepsilon_0}\int_{z_\sigma}^z\sigma_\textrm{EF}^{(+)}\,\dd z'\nonumber\\
&&+V_{z_1}+\phi\\
\sigma(z<-z_1)&=&\int_{-\infty}^z\rho(z')\,\dd z'\nonumber\\
\sigma(z>z_1)&=&\int_{z_1}^z\rho(z')\,\dd z
\end{eqnarray}

\noindent where $V_{z_1}$ is the potential at $z_1$ and $\phi$ is a phase due to the dipole moment, which is only present if the system has neither mirror nor inversion symmetry. The potential is given by

\begin{eqnarray}
V_{z_1} &=& V(-z_1)+\frac{1}{\varepsilon_0}\left(\sigma(-z1)D_\mathrm{vac}+{\textstyle\frac12}D_\mathrm{vac}^2(-\bar\rho)\right)\nonumber\\
\phi &=& -\frac{1}{\varepsilon_0}\sum_{\Gperp\ne0}\int_{-z_1}^{z_1} z\rho_{0,\Gperp}\ee^{\ii\Gperp z}\,\dd z\nonumber\\
&=& -\frac{2\ii}{\varepsilon_0}\sum_{\Gperp\ne0} j_1(\Gperp z_1)\rho_{0,\Gperp}z_1^2,
\end{eqnarray}

\noindent where $j_1(z)=(\sin z-z\cos z)/z^2$.

For the $\Gpar\ne0$ component the solution is given by Eq.~(\ref{eq:gne0:V}); for the FLAPW basis, the integration has to be split into two vacua and the film region; for a given $z>|z_1|$ the integral in the vacuum can be further split into $\int_{z_1}^{z}$ and $\int_z^{z_\sigma}$, which has been done to obtain the functions $\alpha$ and $\beta$ below; the vacuum potential is then given by

\begin{eqnarray}
V_\Gpar(\pm|z|) &=& -\frac{1}{2\varepsilon_0 |\Gpar|}\Bigl[ \ee^{\Gpar|z|} \alpha_\mathbf{G}(\mp|z|)\nonumber\\
&&\qquad+\ee^{-|\Gpar|\cdot|z|} \beta_\Gpar(\mp|z|)\nonumber\\
&&\qquad+\ee^{-|\Gpar|\cdot|z|} \alpha_\Gpar(\pm|z|)\Bigr]
\end{eqnarray}

\noindent with

\begin{eqnarray}
\alpha_\Gpar(\pm |z|) &=& \pm\int_{\pm z}^{\pm\infty} \tilde\rho_\Gpar(z')\ee^{-|\Gpar|\cdot|z'|}\,\dd z'\\
\beta_\Gpar(\pm|z|)   &=& \pm\int_{\pm z_1}^z \tilde\rho_\Gpar(z')\ee^{|\Gpar|\cdot|z'|}\,\dd z',
\end{eqnarray}

\noindent where $\tilde\rho_\Gpar(z)=\rho_\Gpar(z)+\delta(|z|-z_\sigma)\sigma_{\mathrm{EF},\Gpar}^{(\mathrm{sgn}\>z)}$ and $\sigma_{\mathrm{EF},\Gpar}^{(\mathrm{sgn}\>z)}$ is the (inhomogeneous) surface charge density of, respectively, the top and bottom charge sheet.

\subsection{\label{sect:Appendix:Dir}Dirichlet boundary conditions}

Contrary to the Neumann boundary condition used above, which defines the boundary via a surface charge density, the Dirichlet boundary condition has a fixed potential at the boundaries, which matches metallic plates held a certain voltage. The Dirichlet boundary conditions not only imply a different single-particle potential but due to the density-dependent image charges the effective Coulomb interaction between electrons is modified.\cite{Hallam1996,Indlekofer2005} The effect of the latter is not included in the described scheme and would require a modified functional.

For the constant-potential boundary condition, the $\Gpar=0$ part, solved by integrating Eq.~(\ref{eq:poisson}), is given by

\begin{eqnarray}
V_0(z) &=& -\frac{1}{\varepsilon_0}\int_{z_0}^z\int_{z_0}^{z'}\rho_0(z'')\,\dd z''\dd z'\nonumber\\
        && + \frac{1}{\varepsilon_0}(z-z_0)\sigma_0+V_0\nonumber\\
       &=& -\frac{1}{\varepsilon_0}\int_{z_0}^z \sigma(z')\,\dd z' + \frac{1}{\varepsilon_0}(z-z_0)\sigma_0+V_0,\qquad
\end{eqnarray}

\noindent with $\sigma(z) = \int_{z_0}^z\rho(z')\,\dd z'$. The lower boundary is set to $-z_\sigma$, the location of the metallic plate with the associated potential $V_0=V_{-z_\sigma}$. For the other plate at $+z_\sigma$, the potential $V_{z_\sigma}$ is given by

\begin{eqnarray}
V_0(z_\sigma) &=& -\frac{1}{\varepsilon_0}\int_{-z_\sigma}^{z_\sigma}\sigma(z')\,\dd z' + 2z_\sigma \frac{\sigma_0}{\varepsilon_0}+V_0;
\end{eqnarray}

\noindent thus,
\begin{eqnarray}
\frac{\sigma_0}{\varepsilon_0} = \frac{1}{2z_\sigma}\left[V_{z_\sigma}-V_{-z_\sigma} + \frac{1}{\varepsilon_0}\int_{-z_\sigma}^{z_\sigma}\sigma(z')\,\dd z' \right]
\end{eqnarray}

\noindent which can be used in a similar way to the Neumann solution described above.

To solve the Poisson equation for $\Gpar\ne0$, we start with the solution of the homogeneous problem $\left[\partial^2_z-\Gpar^2\right]V_\Gpar(z)=0$, which is given by

\begin{eqnarray}
\tilde V_\Gpar(z) = C_1\ee^{-|\Gpar| z}+C_2\ee^{+|\Gpar| z}.
\end{eqnarray}

The coefficients $C_1$ and $C_2$ have to be chosen such that the boundary condition $V_\Gpar(-z_\sigma)=V_{-z_\sigma}^{(\Gpar)}$ and $V_\Gpar(z_\sigma)=V_{z_\sigma}^{(\Gpar)}$ are fulfilled; one obtains

\begin{eqnarray}
C_1 = \frac{V_{-z_\sigma}^{(\Gpar)} x - V_{z_\sigma}^{(\Gpar)} y}{x^2-y^2},
C_2 = \frac{V_{z_\sigma}^{(\Gpar)} x - V_{-z_\sigma}^{(\Gpar)} y}{x^2-y^2};\qquad\quad
\end{eqnarray}

\noindent hereby $x=\exp(|\Gpar| z_\sigma)$ and $y=\exp(-|\Gpar| z_\sigma)$.  On the other hand, a Green's function solving Eq.~(\ref{eq:poisson}) is given by\cite{Hallam1996}

\begin{eqnarray}
G(z,z') &=& \frac{\displaystyle 1}{\displaystyle |\Gpar|\sinh (\Gpar2z_\sigma)}\\
&&\times \begin{cases}
\sinh \Gpar(z_\sigma-z')\,\sinh \Gpar(z+z_\sigma), & z \le z'\\
\sinh \Gpar(z'+z_\sigma)\,\sinh \Gpar(z_\sigma-z), & z > z'\\
\end{cases}\nonumber
\end{eqnarray}

\noindent which vanishes at the boundary $\pm z_\sigma$; for $z_\sigma\to\infty$ it
simplifies to Eq.~(\ref{eq:GreenF:Neumann}). Using this Green's function, the particular solution is given by

\begin{eqnarray}
V_\Gpar^\textrm{(part)}(z) = \frac{1}{\varepsilon_0}\int G(z,z')\rho_\Gpar(z')\,\dd z'.
\end{eqnarray}

\noindent Combining the regular solution of the homogeneous system with the particular solution gives

\begin{eqnarray}
V_\Gpar (z) &=& \frac{1}{\varepsilon_0}\int_{-z_\sigma}^{z_\sigma} \rho_\Gpar(z')G(z, z')\,\dd z'\nonumber\\
             && + C_1\ee^{-|\Gpar| z}+C_2\ee^{+|\Gpar| z}
\end{eqnarray}

\noindent with the $C_1$ and $C_2$ as defined above.



%

\end{document}